\def\changesHilighted{false} 

\documentclass[lettersize,journal]{IEEEtran}
\usepackage{amsmath,amsfonts}
\usepackage{url}
\usepackage{graphicx}
\usepackage{cite}
\usepackage{amsmath,amssymb}
\usepackage{color}
\usepackage[dvipsnames]{xcolor}
\usepackage{accents}

\DeclareMathOperator*{\argmax}{arg\,max}

\newcommand{\ubar}[1]{\underaccent{\bar}{#1}}

\definecolor{azure}{rgb}{0.0,0.5,0.9}
\definecolor{forestgreen}{rgb}{0.33,0.61,0.34}
\newcommand{\aiyi}[1]{ \textcolor{azure}{\bf [Aiyi: #1]}}
\newcommand{\aiyii}[1]{ \textcolor{azure}{\bf [Aiyi: #1]}}
\newcommand{\masaki}[1]{ \textcolor{Bittersweet}{\bf [Masaki: #1]}}
\newcommand{\masakii}[1]{ \textcolor{Bittersweet}{\bf [Masaki: #1]}}
\newcommand{\us}[1]{ \textcolor{Bittersweet}{\bf [Reply: #1]}}
\newcommand{\waka}[1]{ \textcolor{ForestGreen}{\bf [Prof. Wakamiya: #1]}}

\usepackage[normalem]{ulem}
\newcommand{\add}[1]{\textcolor{blue}{#1}}
\newcommand{\addd}[1]{\textcolor{blue}{#1}}
\newcommand{\del}[1]{\textcolor{red}{\sout{#1}}}
\newcommand{\dell}[1]{\textcolor{red}{\sout{#1}}}
\newcommand{\delwhole}[1]{#1}
\newcommand{\dellwhole}[1]{#1}
\renewcommand{\add}[1]{#1}
\renewcommand{\us}[1]{}
\renewcommand{\waka}[1]{}
\renewcommand{\aiyi}[1]{}
\renewcommand{\del}[1]{}
\renewcommand{\delwhole}[1]{}
\renewcommand{\masaki}[1]{}

\ifnum\pdfstrcmp{\changesHilighted}{false}=0
\renewcommand{\addd}[1]{#1}
\renewcommand{\masakii}[1]{}
\renewcommand{\dell}[1]{}
\renewcommand{\dellwhole}[1]{}
\renewcommand{\aiyii}[1]{}
\fi

\usepackage{amsthm}
\theoremstyle{definition}

\begin{document}
\allowdisplaybreaks
\title{Communication-Free Shepherding Navigation with Multiple Steering Agents}

\author{Aiyi Li, Masaki Ogura,~\IEEEmembership{Member,~IEEE}, Naoki Wakamiya,~\IEEEmembership{Member,~IEEE}
}

\markboth{Journal of \LaTeX\ Class Files,~Vol.~14, No.~8, August~2021}%
{Shell \MakeLowercase{\textit{et al.}}: A Sample Article Using IEEEtran.cls for IEEE Journals}

\IEEEpubid{0000--0000/00\$00.00~\copyright~2021 IEEE}

\maketitle

\begin{abstract}
Swarm guidance addresses a challenging problem considering the navigation and control of a group of passive agents. To solve this problem, shepherding offers a bio-inspired technique of navigating such group of agents by using external steering agents with appropriately designed movement law. Although most shepherding researches are mainly based on the availability of centralized instructions, these assumptions are not realistic enough to solve some emerging application problems. Therefore, this paper presents a decentralized shepherding method where each steering agent makes movements based on its own observation without any inter-agent communication. Our numerical simulations confirm the effectiveness of the proposed method by showing its high success rate and low costs in various placement patterns. These advantages particularly improve with the increase in the number of steering agents.
\end{abstract}


\section{Introduction}

\IEEEPARstart{S}{hepherding} problem~\cite{long2020} refers to the problem of designing the movement law of steering agents (called shepherds) to navigate another set of agents (called sheep) driven by the repulsive force from steering agents, and has been attracting emerging attentions by its applicability in robotics~\cite{chung2018survey}, group dynamics~\cite{vemula2018social}, and 
nanochemistry~\cite{mou2019active}. Specifically, several works have been published toward providing effective solutions to the shepherding problem within various scientific fields including the systems and control theory~\cite{bacon2011,Pierson2018}, robotics~\cite{zhi2020learning}, and the complexity science~\cite{el2020limits}. 

One of the major challenges in the shepherding problem is clarifying how to coordinate \emph{multiple} steering agents for effective guidance. Various methodologies and algorithms have been proposed in the literature. For example, Lien et al.~\cite{Lien2005} have illustrated the effectiveness of the navigation by steering agents taking a prescribed formation. Extending this work, Pierson and Schwager~\cite{Pierson2018} have presented a 3-D herding algorithm based on the dimension reduction of the whole multi-agent system. Similar works can be found in~\cite{song2021herding} and~\cite{Chipade2020adversarial}, where caging-based algorithms for guiding a flock of agents are proposed. El-Fiqi et al.~\cite{el2020limits} have presented a centralized shepherding algorithm that assigns a path to each steering agent. Bacon and Olgac~\cite{bacon2011} have presented a quasi-decentralized control law for guiding agents with multiple steering agents based on sliding mode control. 

\IEEEpubidadjcol

Although most of the existing shepherding algorithms with multiple steering agents assume the existence of a central coordinator~\cite{Lien2005,Pierson2018,el2020limits,song2021herding,Chipade2020adversarial,bacon2011}, this assumption requires the coordinator's ability of observing the whole system and the steering agents' ability of communication. However, these requirements can severely limit the practical feasibility of the algorithms.
Although we can find in the literature a few decentralized shepherding algorithms with multiple steering agents~\cite{lee2017autonomous,Hu2020}, these works still implicitly assume the communication among steering agents. 

\IEEEpubidadjcol

The objective of this paper is to propose 
an algorithm for communication-free shepherding navigation with multiple steering agents. Our approach is to start from an existing single-shepherd algorithm called Farthest-Agent Targeting algorithm~\cite{tsunoda2018analysis}. Leveraging on the simplicity of the algorithm, we then construct an algorithm for the shepherding by multiple steering agents 
under the assumption that each shepherd knows \del{only its own relative position to the goal and to other agents within its}
\add{its relative position to the goal, and the relative position of other agents within the shepherd's} recognition range\add{~to the goal}\waka{This does not necessarily mean that a shepherd knows the distance of each agents (sheep) to the goal. I mean, the assumption is not sufficient to realize FAT.}\us{We agree that this was still not clear. For clarity, we have chosen to use another (but equivalent) assumption on the information available to the shepherd.}%
. Within the proposed algorithm, although each shepherd attempts to guide the whole flock by chasing its own target sheep independently and without inter-shepherd communication, cooperative behavior emerges as a consequence of spatial distribution of the shepherds induced by the inter-shepherd repulsion built into the algorithm. A shepherd's target sheep is determined as the sheep maximizing the weighted difference of the sheep's distance from the goal and the one from the shepherd. The effectiveness of the proposed algorithm is illustrated with extensive numerical simulations. 

\IEEEpubidadjcol

This paper is organized as follows. In Section~\ref{sec:prob}, we state the problem studied in this paper. In Section~\ref{sec:prop}, we describe our communication-free shepherding algorithm. We finally present the numerical simulations in Section~\ref{sec:num}.

\section{Problem Statement} \label{sec:prob}

We consider the situation where there exist $N$ agents to be navigated, called sheep, and~$M$ steering agents, called shepherd. Throughout this paper, we use the notations $[N] = \{1, 2, \dotsc, N\}$ and~$[M] = \{1,2, \dotsc, M\}$. The sheep and shepherd are assumed to dynamically move on the two-dimensional space~$\mathbb{R}^{2}$ in the discrete-time. \addd{For any $i\in [N]$, $k\in [M]$, and $t = 0, 1, 2, \dotsc$, we let $p_i(t)$ ($q_k(t)$) denote the position of the $i$th sheep ($k$th shepherd, respectively) at time~$t$.}\masakii{Made a few minor modifications.} We suppose that each sheep has a limited range~$r>0$ of recognizing other agents. Therefore, the indices of the sheep and the shepherd agents that can be recognized by the $i$th sheep at time~$t$ are given by 
  $\mathcal N_i(t) 
= \{ j \in [N] \mid 
     \addd{0< \lVert p_i(t) - p_j(t) \rVert < r}
    \}$
    and~$
    \mathcal M_i(t) 
    = \{ \ell \in [M] \mid 
     \addd{0<\lVert p_i(t)-q_\ell(t) \rVert < r}
    \}
    $\masakii{I think it is necessary to exclude $0$ (i.e., strict inequality). Is this correct?}\aiyii{This is correct.}.

As in the literature (see, e.g., \cite{tsunoda2018analysis,strombom2014solving}), we adopt the following equation as the mathematical model for describing the movement of the sheep agents: 
\begin{equation*}
	p_i(t+1) = p_i(t) + u_i(t), 
\end{equation*}
where $u_i(t)\in \mathbb R^2$ represents the movement vector of the $i$th sheep and is supposed to be of the form 
	$u_{i}(t) = {c_1}u_{i1}(t) + {c_2}u_{i2}(t) + {c_3}u_{i3}(t) + {c_4}u_{i4}(t)$. 
In this equation, the vectors~$u_{i1}(t)$, $u_{i2}(t)$, and~$u_{i3}(t) \in \mathbb{R}^2$ are the forces of separation, alignment, and cohesion given by 
\begin{equation}
\begin{aligned}
\label{separation_shp}
	u_{i1}(t) &= -{\lvert \mathcal N_i(t) \rvert }^{-1} \sum_{j\in \mathcal N_i(t)} \psi(p_j(t)-p_i(t)), 
	\\
	u_{i2}(t) &= -{\lvert \mathcal N_i(t) \rvert }^{-1} \sum_{j\in \mathcal N_i(t)} \phi(u_j(t-1)), 
	\\
	u_{i3}(t) &= {\lvert \mathcal N_i(t) \rvert }^{-1} \sum_{j\in \mathcal N_i(t)}
	\phi(p_j(t)-p_i(t))
\end{aligned}
\end{equation}
where $\phi(x) = x/\lVert x\rVert$ is a normalization operator and $\psi(x) = x/\lVert x\rVert^3$ is a potential-like function. We extend the domain of these mappings to the whole space~$\mathbb R^2$ by letting $\phi(0) = 0$ and $\psi(0) = 0$.
Also, the vector~$u_{i4}(t)$ is the force of repulsion from shepherd agents given by 
	$u_{i4}(t) = - {\lvert \mathcal M_i(t)\rvert }^{-1} \sum_{\ell\in \mathcal M_i(t)} 
	\psi(q_\ell(t)-p_i(t))$. 
We remark that, when the set~$\mathcal{N}_i(t)$ is empty, we regard the vectors in~\eqref{separation_shp} to be the zero vectors. Likewise, if the set~$\mathcal{M}_i(t)$ is empty, then we set $u_{i4}(t) = 0$. This rule applies to other similar equations appearing later in this paper. 

As for the shepherd agents, we place the following restriction on the information available for navigation. The recognition range of a shepherd is assumed to be finite and is set to be $r'>0$; therefore, the set of indices of the sheep and the shepherd agents that can be recognized by the $k$th shepherd at time~$t$ are given by 
    $\mathcal N'_k(t) = \{ j \in [N] \mid 
     \addd{0< \lVert q_k(t)-p_j(t) \rVert < r'}
    \}$ and 
    $\mathcal M'_k(t) = \{ \ell \in [M] \mid 
     \addd{0<\lVert q_k(t)-q_\ell(t) \rVert < r'}
    \}$\masakii{Same as above.}.

We further assume that a shepherd can detect only \del{its relative position to the goal and to other agents within its own}\add{its relative position to the goal, and the relative position of other agents within the shepherd's} recognition range\add{~to the goal}. \dell{This implies that}\addd{Hence,}\masakii{Just to shorten this paragraph/} the $k$th shepherd needs to determine its own movement at time~$t$ based only on the following types of the vectors: 
\begin{itemize}
    \item \del{$p_j(t)-q_k(t)$ and }$p_j(t)-x_g$ (\,$j \in \mathcal N'_k(t)$);
    \item \del{$q_\ell(t)-q_k(t)$ and }$q_\ell(t)-x_g$ ($\ell \in \addd{\{k\}\cup \mathcal M'_k(t)}$)\masakii{Took union with the sheep $k$ itself}
\end{itemize}
Finally, the objective of the navigation by shepherd agents is to herd all the sheep into a goal region~$G \subset \mathbb R^2$. In this paper, we suppose that the goal region~$G$ is the closed disk with center~$x^{g}\in \mathbb R^2$ and radius~$R^{g}> 0$.

\section{Proposed Algorithm}\label{sec:prop}

In this section, we describe the algorithm we propose for the movement of the shepherd agents. We start by recalling the Farthest-Agent Targeting (FAT) algorithm~\cite{tsunoda2018analysis} designed for the case of a single shepherd (i.e., $M=1$). In the algorithm, the movement of the (1st) shepherd is specified as $q_1(t+1) = q_1(t) + v_1(t)$, where $v_1(t) \in \mathbb{R}^2$ represents the movement vector of the shepherd. Let us denote the position of the sheep agent farthest from the goal by $\xi_1(t)$; i.e., define $\xi_1(t) = \argmax_{p \in \{p_i(t)\}_{i \in [N]}} {\lVert p-x_g \rVert}$. Then, in the FAT algorithm, the movement vector~$v_1(t)$ is specified as the weighted sum of the following three vectors:
\begin{align}
	\phi(\xi_1(t)-q_1(t)),\ 
	-\psi(\xi_1(t)-q_1(t)),\ 
    -\phi(x_g-q_1(t)), 
    \label{eq:towardgoal}
\end{align}
which are, respectively, to realize the movement of the shepherd for chasing the farthest agent, taking an appropriate distance with the farthest agent, and pushing the farthest agent toward the goal region. 

Despite being simple, the FAT algorithm is known for its effectiveness in performing the shepherding navigation with a single shepherd~\cite{tsunoda2018analysis}. However, the algorithm requires a global knowledge of the positions of all sheep. Furthermore, when generalized to the situation of multiple shepherds, the formula would result in all the shepherds targeting the same sheep, which 
is presumably inefficient. We have also found through our preliminary evaluation that the FAT algorithm tends to scatter the flock when another sheep lies between the shepherd and the farthest sheep. 

Based on these observations, in this paper,
we propose an extended version of the FAT algorithm to let each shepherd choose, as its target, a sheep both close to itself and far from the goal. Specifically, we propose that the sheep agent targeted by the $k$th shepherd is determined by the formula
\begin{equation}
  \label{nearer}
  \xi_k(t) = 
  \argmax_{p \in \{p_{j}(t)\}_{j\in \mathcal N'_k(t)}} 
  \bigl(
  \lVert p-x_g \rVert 
  - 
  \alpha \lVert p-q_k(t) \rVert
  \bigr)
\end{equation}
for a constant~$\alpha > 0$. We remark that $\xi_k(t)$ is decidable by the $k$th shepherd because the sheep's relative position to the \dell{goal}{\addd{shepherd}} is computable as \del{$p_j(t)-x_g = (p_j(t)-q_k(t)) - (q_k(t)-x_g)$}\add{$p_j(t)-q_k(t) = (p_j(t)-x_g) - (q_k(t)-x_g)$}\aiyi{I change the sequence to be consistent with Formula~\ref{nearer}.}. We also remark that the additional term $\alpha \lVert p-q_k(t) \rVert$ in \eqref{nearer} is introduced to address the second and the third issues of the FAT algorithm. 
\begin{figure*}
\centering
\includegraphics[width=.78\linewidth]{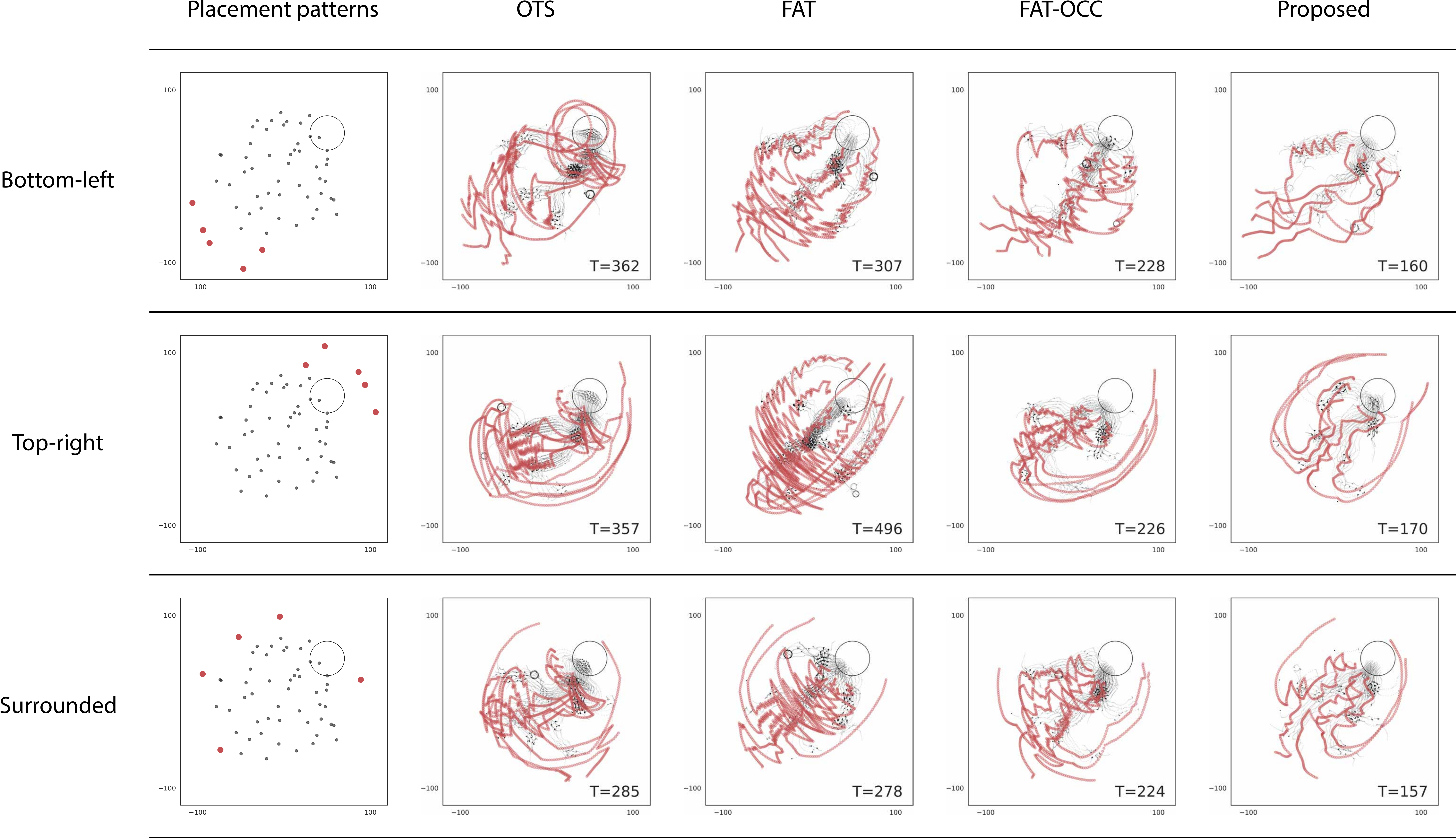}
  \caption{1st column: Samples of initial configurations.\us{We select the samples that are different from all the trials. These three samples have the same distribution of sheep flock} 2nd to 5th columns: trajectories of the quickest navigations among those performed for randomly generated 100 initial configurations. Circle: goal region. Red dots: shepherds. Gray dots: sheep. The numbers at the bottom-right indicate the time at which the shepherding navigation is completed. It is remarked that the initial configurations in each row are not necessarily the same.}
  \label{fig:trace}
\end{figure*}

We can now state the proposed movement algorithm of the shepherds. As in the FAT algorithm, we let
\begin{equation*}
	q_k(t+1) = q_k(t) + v_k(t), 
\end{equation*}
where $v_k(t)$ denotes the movement vector of the $k$th shepherd. This vector is to be constructed as the weighted sum of the following four vectors. First, we define 
	$
	v_{k1}(t) =
	\phi
	(\xi_k(t)-q_k(t) 
)
	$
for the $k$th shepherd to chase the target sheep. Secondly, in order to take an appropriate distance between the shepherd and sheep agents, we define
\begin{equation}
  v_{k2}(t) = - {\lvert \mathcal N'_k(t)\rvert }^{-1} \sum_{j\in \mathcal N'_k(t)}
  \psi\bigl(p_j(t)-q_k(t)\bigr). 
\label{eq:vk2}
\end{equation}
Thirdly, to achieve the guidance toward the goal region, we define the vector 
$
v_{k3}(t) = -
\phi
(x_g-q_k(t)
)
$\dell{.} by adopting \eqref{eq:towardgoal}. Finally, in order to avoid competition among shepherd agents for efficient guidance, we introduce the vector
\addd{$
	v_{k4}(t) = - {\lVert x_g-q_k(t) \rVert}\,\lvert \mathcal M'_k(t)\rvert ^{-1} \sum_{\ell\in \mathcal M'_k(t)}
	\psi(q_\ell(t)-q_k(t))$,} 
\dellwhole{\begin{equation*}
	\dell{v_{k4}(t) = - {\lVert x_g-q_k(t) \rVert}\,\lvert \mathcal M'_k(t)\rvert ^{-1} \sum_{\ell\in \mathcal M'_k(t)}
	\psi\bigl(q_\ell(t)-q_k(t)\bigr),} 
\end{equation*}}
which represents repulsion between shepherd agents. Because shepherds needs to be relatively closer to each other at the final stage of the shepherding navigation, we are introducing the weight term~$\lVert x_g-q_k(t) \rVert$. \add{The relative position between shepherd agents are also computable as $q_{\dell{l}\addd{\ell}}(t)-q_k(t) = (q_{\dell{l}\addd{\ell}}(t)-x_g) - (q_k(t)-x_g)$.} Now, based on the four vectors introduced above, we define the movement vector of the $k$th shepherd as 
	$
	v_{k}(t) = d_1v_{k1}(t) + d_2v_{k2}(t) + d_3v_{k3}(t) + d_4v_{k4}(t)
	$
for positive constants $d_1$, $d_2$, $d_3$, and $d_4$.


\section{Numerical Simulation}\label{sec:num}

In this section, we present numerical simulations to illustrate the effectiveness of the proposed algorithm.

\subsection{Configuration}

We assume that there exist $N=50$ sheep to be guided. They are placed uniformly and randomly on the disc centered at the origin and having radius~$80$. Their parameters are set as $c_1=100$, $c_2 = 0.5$, $c_3 = 2$, $c_4 = 400$, and~$r = 20$. The goal~$G$ is supposed to have the center~$x^g = [50, 50]^\top$ and radius~$R^g = 20$. For the comprehensiveness of our experiment, we prepare the following three different placement patterns of the shepherd agents; shepherd agents are initially 1) placed around at the bottom-left of the sheep agents (\emph{bottom-left}), 2) placed around at the top-right of the sheep agents (\emph{top-right}), and 3) surrounding the sheep agents (\emph{surrounding}). For each of the placement patterns, we randomly generate $100$ different initial configurations of agents. Samples of the initial configurations are shown in the first column of Fig.~\ref{fig:trace}. 

Furthermore, in the numerical simulations, we modify the operator~$\psi$ in the original model as 
\begin{equation*}
    \psi(x) = \begin{cases}
      x/\lVert x\rVert^3, &\mbox{if $\lVert x\rVert \geq \ubar r$,}
      \\
      x/(\lVert x \rVert \ubar r^2), &\mbox{if $0 < \lVert x\rVert < \ubar r$,}
      \\
      0,&\mbox{otherwise}
    \end{cases}
\end{equation*}
to avoid the numerical instability of the operator~$\psi$ in its original definition.
We use $\ubar r = 3$, which allows us to avoid the numerical instability while ensuring that the separation force is still dominant between agents close enough to each other.

\subsection{Proposed and baseline algorithms}

We describe the algorithms to be compared in our numerical simulations. For each of the initial configuration, all of the algorithms are to be terminated when all the sheep agents belong to the goal~$G$, or after $3000$ steps regardless of the result of navigation. In the former case, we label the trial for the initial configuration as success. 

\subsubsection{Proposed algorithm}

We use the parameters $d_1 = 2.5$, $d_2 = 100$, $d_3= 1$, $d_4 = 2$, and~$r'=100$. The parameter~$\alpha$ in equation~\eqref{nearer} is set as $\alpha = 1$. 

\subsubsection{Farthest-Agent Targeting}

When $\alpha = 0$, the formula~\eqref{nearer} lets each shepherd target the sheep farthest from the goal among all the sheep in the shepherd's recognition range
. For this reason, we call the proposed algorithm with these specific parameters as the Farthest-Agent Targeting (FAT) algorithm.

\subsubsection{Farthest-Agent Targeting with Occlusion} 

The farthest-agent targeting algorithm with occlusion (FAT-OCC) \cite{tsunoda2018analysis} is also considered. This algorithm is identical to the FAT algorithm except that the vector~$v_{k2}(t)$ in \eqref{eq:vk2} is modified as 
  $v_{k2}(t) = - \lvert \mathcal N'_{k, \textrm{occ}}(t)\rvert^{-1} \sum_{j\in \mathcal N'_{k, \textrm{occ}}(t)}
  \psi(p_j(t)-q_k(t))$, 
in which the set~$\mathcal N'_{k, \textrm{occ}}(t)$ represents the set of sheep agents recognizable under occlusion and is constructed as follows. For each $t$, we first initialize $\mathcal N_{k, \textrm{occ}}'(t) =\emptyset$. We then order the set~$\mathcal N_k'(t)$ as $(i_1, \dotsc, i_{\lvert \mathcal N_k'(t)\rvert})$ in such a way that $\lVert p_{i_1}(t) \rVert \leq \lVert p_{i_2}(t) \rVert \leq \cdots \leq \lVert p_{i_{\lvert \mathcal N_k'(t)\rvert }}(t) \rVert$. For each $\iota = 1, \dotsc, \lvert \mathcal N_k'(t)\rvert$, we sequentially join the index~$i_{\iota}$ to the set~$\mathcal N_{k, \textrm{occ}}'(t)$ if and only if $\lvert \arctan (p_\iota-q_k) - \arctan{(p_\phi-q_k)}\rvert > \theta$ for all $\phi \in \mathcal N_{k, \textrm{occ}}'(t)$. We use the parameter~$\theta = 0.05$. 

\subsubsection{Online-Target Switching}

The Online-Target Switching (OTS) algorithm proposed by Str\"ombom~\cite{strombom2014solving} is applied by judging the flock separation. We implement this algorithm by replacing $\xi_k(t)$ in \eqref{nearer} by $\xi^{\textrm{ots}}_k(t)$ defined by 
\begin{equation*}\displaystyle
  \label{switching_offset}
  \xi^{\textrm{ots}}_k(t) = 
  \begin{cases}
    \bar{p}(t) + d\,{\phi\bigl(\bar{p}(t)-x_g\bigr)},\!\!\!\!\!\!\!\!\!\!\!\!
    &\mbox{\!\!\!\!\!if $\lVert p^{\#}(t)-\bar p(t) \rVert\leq R^{\textrm{ots}}$},
    \\
	\bar p(t) + d\,{\phi\bigl(p^{\#}(t)-\bar{p}(t)\bigr)},\!\!\!\!\!
	&\mbox{otherwise}
  \end{cases}
\end{equation*}
where $\bar p(t) = N^{-1} \sum_{i=1}^N p_i(t)$ is the mass center of the sheep flock and~$p^{\#}(t) = \argmax_{p\in \{p_i(t)\}_{i \in [N]}}{\lVert \bar p(t)-p \rVert}$ represents the position of the sheep farthest from the mass center. We use $R^{\textrm{ots}}=25$ and~$d^{\textrm{ots}} = 4$. 

\subsection{Simulation Results}

We first illustrate the behaviors of the algorithms using the trajectories of agents. Toward this end, for each pair of the four algorithms and three placement patterns, we pick the quickest trial among 100 initial configurations. The trajectories and their corresponding completion time are shown in Fig.~\ref{fig:trace}. We can observe that the trajectories of the shepherds in the proposed algorithm are smoother than those of the three baseline algorithms, confirming the effectiveness of the decentralized mechanism of the proposed algorithm.

For further evaluation and comparison of the proposed and the baseline algorithms, we introduce the following three performance measures. First, the success rate of an algorithm for a placement pattern is defined as the rate of successful trials among randomly generated $100$ initial configurations. Second, we define the completion time as the execution time of the algorithm in its successful trials. Finally, the average path length is defined as the average of the mean traveling distance $M^{-1}\sum_{k=1}^M\sum_{t}\lVert v_k(t)\rVert$ of shepherds in successful trials. 

\begin{figure}[tb]
  \centering
  \includegraphics[width=0.99\linewidth]{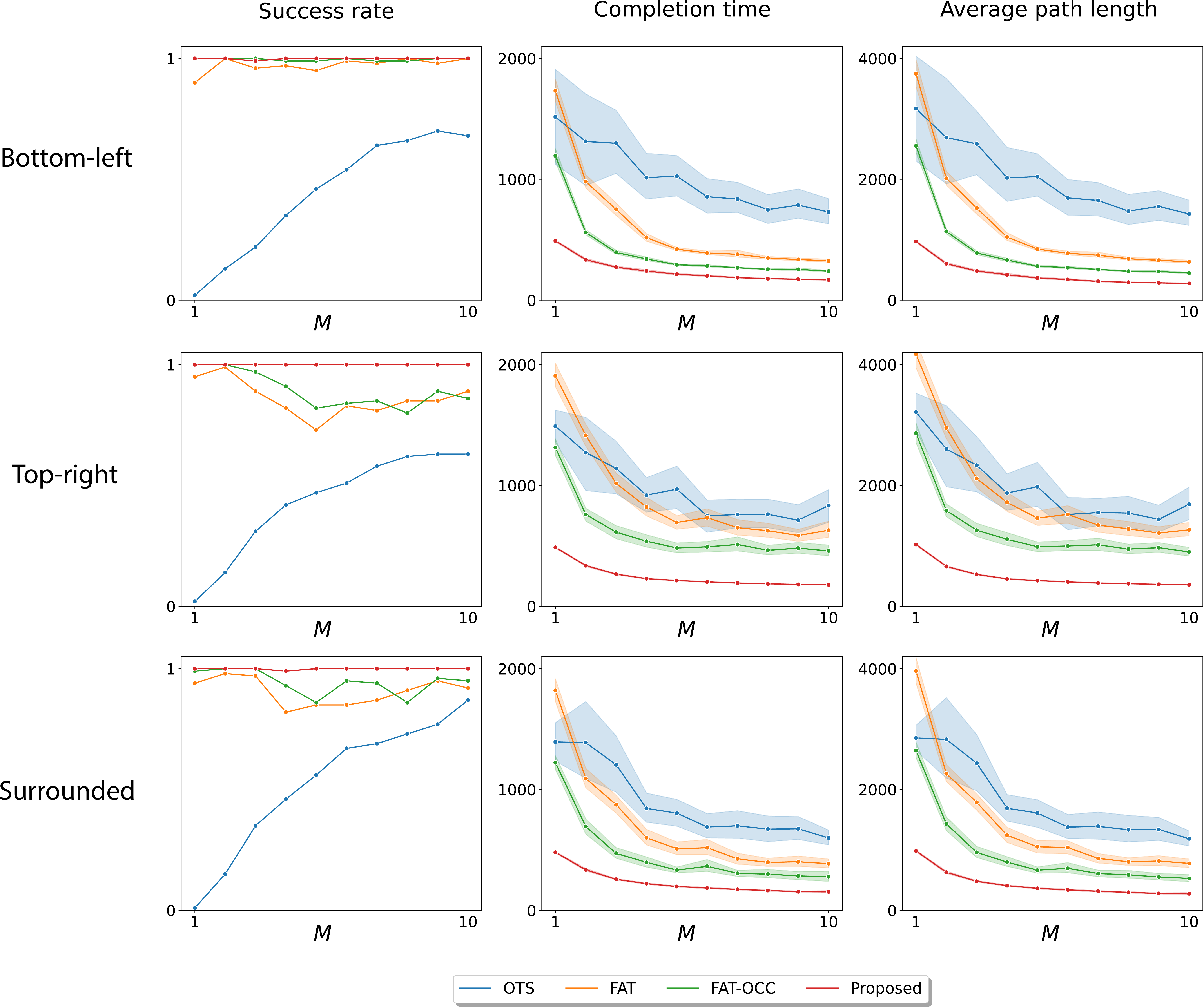}
\caption{Performances of the four algorithms. Horizontal axes represent the number of shepherds. 1st column: the rate of successful navigation. 2nd column: success time. 3rd column: average traversal distance of shepherds. In 2nd and 3rd columns, a solid line draws an estimate of the mean value and shaded areas describe confidence interval for that estimate.}
\label{fig:analysis}
\end{figure}

Fig.~\ref{fig:analysis} shows how these three performance measures depend on the number of shepherds for each of the algorithms. We observe that the proposed algorithm achieves almost 100\% success rate regardless of the number of the shepherds and placement patterns, which confirms the effectiveness and robustness of the proposed algorithm. We can also observe that the proposed algorithm outperforms the baseline methods in both completion time and average path length. Furthermore, the average completion time and average path length steadily decreases with respect to the number of shepherds. These trends suggest that the proposed algorithm allows stable and synergistic coordination of shepherds for the navigation of sheep agents.  

\section{Conclusion}


In this paper, we have studied the shepherding problem with multiple steering agents unable to communicate with each other. We have first presented a model of sheep agents in the presence of multiple steering agents. We have then proposed a distributed and communication-free
algorithm with multiple steering agents to aggregate the sheep agents by a location-based self planning. Finally, we have confirmed the effectiveness and robustness of the proposed algorithm via extensive numerical simulations. Interesting directions of future works include investigating if the proposed communication-free coordination mechanism can be extended to other types of navigation tasks.

\end{document}